\documentclass[11pt,a4paper,dvips]{article}\usepackage[latin1]{inputenc}
\usepackage{graphicx}
\newcommand{\qed}{\hspace*{\fill}$\square$}
\newtheorem{theorem}{Theorem}
\newtheorem{corollary}[theorem]{Corollary}

\newtheorem{lemma}[theorem]{Lemma}

\usepackage{amsmath,amsfonts,amscd,amssymb}
\begin{document}
\title{%{\small antoine/article/discriminant/NcubeARTICLE$\{$.tex, .aux, .log, .ps, .dvi$\}$}\\
Discriminating and Identifying Codes\\
in the Binary Hamming Space}
\author{
           {\bf Ir\`ene Charon}\\
           GET - T\'el\'ecom Paris \& CNRS - LTCI UMR 5141\\
           46, rue Barrault, 75634 Paris Cedex 13 - France\\
           irene.charon@enst.fr
\and
           {\bf G\'erard Cohen}\\
           GET - T\'el\'ecom Paris \& CNRS - LTCI UMR 5141\\
           46, rue Barrault, 75634 Paris Cedex 13 - France\\
           gerard.cohen@enst.fr
\and
           {\bf Olivier Hudry}\\
           GET - T\'el\'ecom Paris \& CNRS - LTCI UMR 5141\\
           46, rue Barrault, 75634 Paris Cedex 13 - France\\
           olivier.hudry@enst.fr
\and
           {\bf Antoine Lobstein}\\
           CNRS - LTCI UMR 5141 \& GET - T\'el\'ecom Paris\\
           46, rue Barrault, 75634 Paris Cedex 13 - France\\
           antoine.lobstein@enst.fr\\   
$\:$\\}
%\date{}
\date{}
\maketitle
\begin{abstract}Let $F^n$ be the binary $n$-cube, or binary Hamming space of dimension~$n$, endowed with the Hamming distance, and ${\cal E}^n$ (respectively, ${\cal O}^n$) the set of vectors with even (respectively, odd) weight. For $r\geq 1$ and $x\in F^n$, we denote by $B_r(x)$ the ball of radius~$r$ and centre~$x$. A code $C\subseteq F^n$ is said to be $r$-identifying if the sets $B_r(x) \cap C$, $x\in F^n$, are all nonempty and distinct. A code $C\subseteq {\cal E}^n$ is said to be $r$-discriminating if the sets $B_r(x) \cap C$, $x\in {\cal O}^n$, are all nonempty and distinct. We show that the two definitions, which were given for general graphs, %(pas tout a fait vrai)
are equivalent in the case of the Hamming space, in the following sense: for any odd $r$, there is a bijection between the set of $r$-identifying codes in~$F^n$ and the set of $r$-discriminating codes in~$F^{n+1}$. We then extend previous studies on constructive upper bounds for the minimum cardinalities of identifying codes in the Hamming space.
\end{abstract}
\pagebreak

$\mbox{  }$

\medskip

\noindent {\bf Running Head:} 
Identifying codes in Hamming space

\medskip


\medskip

\noindent {\bf Corresponding Author:}\\
G. Cohen, ENST, Dpt INFRES\\
46, rue Barrault, 75634 Paris cedex 13 - France\\
e-mail: cohen@enst.fr $\: \:$ fax: +33 1 45 81 31 19

\bigskip

%$\;$
%
%\bigskip
%
%
%$\;$
%
%\bigskip
%
%$\;$
%
%\bigskip
%
\section{Introduction}
\label{intro}
We define identifying and discriminating codes in a connected, undirected graph $G=(V,E)$, in which a {\it code} is simply a nonempty subset of vertices. These definitions can help, in various meanings, to unambiguously determine a vertex. The motivations may come from processor networks where we wish to locate a faulty vertex under certain conditions, or from the need to identify an individual, given its set of attributes.

In $G$ we define the usual distance $d(v_1,v_2)$ between two vertices $v_1,v_2 \in V$ as the smallest possible number of edges in any path between them. For an integer $r\geq 0$ and a vertex $v\in V$, we define $B_r(v)$ (respectively, $S_r(v)$), the {\it ball} (respectively, {\it sphere}) of radius~$r$ centred at~$v$, as the set of vertices within distance (respectively, at distance exactly) $r$ from~$v$. Whenever two vertices $v_1$ and $v_2$ are such that $v_1 \in B_r(v_2)$ (or, equivalently, $v_2 \in B_r(v_1)$), we say that they $r$-{\it cover} each other. A set $X \subseteq V$ $r$-covers a set $Y \subseteq V$ if every vertex in $Y$ is $r$-covered by at least one vertex in~$X$.

The elements of a code $C\subseteq V$ are called {\it codewords}. For each vertex $v\in V$, we denote by
$$K_{C,r}(v) = C \cap B_r(v)$$
the set of codewords $r$-covering~$v$. Two vertices $v_1$ and $v_2$ with $K_{C,r}(v_1) \neq K_{C,r}(v_2)$ are said to be $r$-{\it separated} by code~$C$, and any codeword belonging to exactly one of the two sets $B_r(v_1)$ and $B_r(v_2)$ is said to $r$-{\it separate} $v_1$ and~$v_2$.

A code $C \subseteq V$ is called $r$-{\it identifying} \cite{karp98} if all the sets $K_{C,r}(v)$, $v\in V$, are nonempty and distinct. In other words, every vertex is $r$-covered by at least one codeword, and every pair of vertices is $r$-separated by at least one codeword. Such codes are also sometimes called {\it differentiating dominating sets}~\cite{gimb01}.

We now suppose that $G$ is bipartite: $G=(V=I\cup A,E)$, with no edges inside $I$ nor~$A$ --- here, $A$ stands for {\it attributes} and $I$ for {\it individuals}. A code $C \subseteq A$ is said to be $r$-{\it discriminating}~\cite{char07} if all the sets $K_{C,r}(i)$, $i \in I$, are nonempty and distinct. {F}rom the definition we see that we can consider only odd values of~$r$. 

In the following, we drop the general case and turn to the binary Hamming space of dimension~$n$, also called the binary $n$-cube, which is a regular bipartite graph. First we need to give some specific definitions and notation.

We consider the $n$-cube as the set of binary row-vectors of length~$n$, and as so, we denote it by $G=(F^n,E)$ with $F=\{0,1\}$ and $E=\{\{x,y\}: d(x,y)=1\}$, the usual graph distance $d(x,y)$ between two vectors $x$ and~$y$ being called here the {\it Hamming distance} --- it simply consists of the number of coordinates where $x$ and $y$ differ. The {\it Hamming weight} %, $w(x)$, 
of a vector~$x$ is its distance to the all-zero vector, i.e., the number of its nonzero coordinates. A vector is said to be {\it even} (respectively, {\it odd}) if its weight is even (respectively, odd), and we denote by ${\cal E}^n$ (respectively,~${\cal O}^n$) the set of the $2^{n-1}$ even (respectively, odd) vectors in~$F^n$. Without loss of generality, for the definition of an $r$-discriminating code, we choose the set $A$ to be ${\cal E}^n$, and the set $I$ to be~${\cal O}^n$. Additions are carried coordinatewise and modulo two.

We denote by $0^n$ (respectively, $1^n$) the all-zero (respectively, all-one) vector of length~$n$. Given a vector $x\in F^n$, we denote by $\pi (x)$ its parity-check bit: $\pi (x)=0$ if $x$ is even, $\pi (x) =1$ if $x$ is odd. Therefore, if~$|$~stands for concatenation of vectors, $x|\pi (x)$ is an even vector. For two sets $X\subseteq F^{n_1}$, $Y\subseteq F^{n_2}$, the {\it direct sum} of $X$ and~$Y$, denoted by $X\oplus Y$, is defined by $X\oplus Y = \{ x|y\in F^{n_1+n_2}: x \in X, y \in Y\}$. %; clearly, for $X,Z \subseteq F^{n_1},Y \subseteq F^{n_2}$,
%$$(X \cup Z) \oplus Y = (X \oplus Y) \cup (Z \oplus Y).$$
Finally, we denote by $M_r(n)$ (respectively,~$D_r(n)$) the smallest possible cardinality of an $r$-identifying (respectively, $r$-discriminating) code in~$F^n$.

In Section~\ref{sec2}, we show that in the particular case of Hamming space, the two notions of $r$-identifying and $r$-discriminating codes actually coincide for all odd values of~$r$ and all $n\geq 2$, in the sense that there is a bijection between the set of $r$-identifying codes in~$F^n$ and the set of $r$-discriminating codes in~$F^{n+1}$. In Section~\ref{sec3}, we give various methods for constructing identifying codes, thus obtaining, in Section~\ref{resu}, upper bounds on~$M_r(n)$, of which several are new. These bounds are summarized in Tables at the end of the paper.
\section{Identifying is discriminating}
\label{sec2}
As we now show with the following two theorems, for any odd $r\geq 1$, any $r$-identifying code in $F^n$ can be extended into an $r$-discriminating code in $F^{n+1}$, and any $r$-discriminating code in $F^{n}$ can be shortened into an $r$-identifying code in $F^{n-1}$. First, observe that $r$-identifying codes exist in~$F^n$ if and only if $r<n$.
\begin{theorem} \label{theo001}
Let $n\geq 2, p\geq 0$ be such that $2p+1<n$, let $C \subseteq F^n$ be a $(2p+1)$-identifying code and let
$$C' = \{c|\pi (c) : c\in C \}.$$
Then $C'$ is $(2p+1)$-discriminating in $F^{n+1}.$ Therefore,
\begin{equation} \label{OH1}
D_{2p+1}(n+1) \leq M_{2p+1}(n).
\end{equation}
\end{theorem}
\noindent {\bf Proof.} Let $r=2p+1$. By construction, $C'$ contains only even vectors. We shall prove that (a) any odd vector $x \in {\cal O}^{n+1}$ is $r$-covered by at least one codeword of~$C'$; (b) given any two distinct odd vectors $x,y \in {\cal O}^{n+1}$, there is at least one codeword in~$C'$ which $r$-separates them.

(a) We write $x=x_1|x_2$ with $x_1\in F^n$ and $x_2\in F$. Because $C$ is $r$-identifying in~$F^n$, there is a codeword $c\in C$ with $d(x_1,c)\leq r$. Let $c'=c|\pi (c)$.

If $d(x_1,c)\leq r-1$, then whatever the values of $x_2$ and $\pi (c)$ are, we have $d(x,c') \leq r$; we assume therefore that $d(x_1,c)=r=2p+1$, which implies that $x_1$ and $c$ have different parities. Since $x_1|x_2$ and $c|\pi (c)$ also have different parities, we have $x_2=\pi (c)$ and $d(x,c')=r.$ So the codeword $c'\in C'$ $r$-covers~$x$.

(b) We write $x=x_1|x_2$, $y=y_1|y_2$, with $x_1, y_1 \in F^n$, $x_2,y_2 \in F$. Since $C$ is $r$-identifying in~$F^n$, there is a codeword $c\in C$ which is, say, within distance~$r$ from $x_1$ and not from~$y_1$: $d(x_1,c)\leq r$, $d(y_1,c)>r$. Let $c'=c|\pi (c)$.

For the same reasons as above, $x$ is within distance $r$ from~$c'$, whereas obviously, $d(y,c')\geq d(y_1,c)>r$. So $c'\in C'$ $r$-separates $x$ and~$y$.

\medskip

Inequality~(\ref{OH1}) follows. \qed
\begin{theorem} \label{theequiv}
Let $n\geq 3, p\geq 0$ be such that $2p+2<n$, let $C \subseteq {\cal E}^n$ be a $(2p+1)$-discriminating code and let
$C' \subseteq F^{n-1}$ be any code obtained by the deletion of one coordinate in~$C$. Then $C'$ is $(2p+1)$-identifying in $F^{n-1}.$ Therefore,
\begin{equation} \label{OH2}
M_{2p+1}(n-1) \leq D_{2p+1}(n).
\end{equation}
\end{theorem}
\noindent {\bf Proof.} Let $r=2p+1$. Let $C \subseteq {\cal E}^n$ be an $r$-discriminating code and $C' \subseteq F^{n-1}$ be the code obtained by deleting, say, the last coordinate in~$C$. We shall prove that (a) any vector $x \in F^{n-1}$ is $r$-covered by at least one codeword of~$C'$; (b) given any two distinct vectors $x,y \in F^{n-1}$, there is at least one codeword in~$C'$ which $r$-separates them.

(a) The vector $x|(\pi (x)+1) \in F^n$ is odd. As such, it is $r$-covered by a codeword $c=c'|u \in C \subseteq {\cal E}^n$: $c' \in C'$, $u=\pi(c')$, and $d(x|(\pi (x)+1), c)\leq r$. This proves that $x$ is within distance $r$ from a codeword of~$C'$.

(b) Both $x|(\pi (x)+1)$ and $y|(\pi (y)+1)$ are odd vectors in $F^n$, and there is a codeword $c=c'|u \in C \subseteq {\cal E}^n$, with $c' \in C'$, $u=\pi(c')$, which $r$-separates them: without loss of generality, $d(x|(\pi (x)+1),c)\leq r$ whereas $d(y|(\pi (y)+1),c)$, which is an odd integer, is at least~$r+2$. Then obviously, $d(x,c')\leq r$ and $d(y,c')\geq r+1$, i.e., there is a codeword in~$C'$ which $r$-separates $x$ and~$y$.

\medskip

Inequality~(\ref{OH2}) follows. \qed
\begin{corollary} \label{coroh}
For all $n\geq 2$ and $p\geq 0$ such that $2p+1<n$, we have: 
$$D_{2p+1}(n+1)=M_{2p+1}(n).$$ \qed
\end{corollary}
It follows that, in the Hamming space, the complexity of problems on discriminating codes is the same as that for identifying codes; in particular, it is known~\cite{honk02b} that deciding whether a given code $C\subseteq F^n$ is $r$-identifying is co-NP-complete.

We now turn to constructions of identifying codes in the $n$-cube, since this is equivalent to our initial goal of constructing discriminating codes.

For previous works, we refer to, e.g., \cite{blas99}--\cite{blas01}, \cite{exoo99}, \cite{honk02a}, \cite{honk02b} or~\cite{karp98}. In the recent~\cite{exoo08}, tables for exact values or bounds on $M_1(n)$, $2\leq n \leq 19$, and $M_2(n)$, $3\leq n \leq 21$, are given.
\section{Constructing identifying codes}
\label{sec3}
We use the notation $(r,n)$ or $(r,n)K$ for a code in $F^n$ which is $r$-identifying and has $K$ elements. Our constructions will use Theorem~\ref{hallucin} below, as well as various heuristics.
\subsection{Extending an identifying code}
\label{sec31}
In the constructions of Theorems~\ref{hallucin} and~\ref{hallucinbis} below, we use a new definition: a code is called $r$-{\it separating} if every pair of vertices is $r$-separated by at least one codeword~\cite[Sec.~3]{blas00} (we do not require anymore that every vertex be $r$-covered by at least one codeword). The following remark and lemma are easy.

\noindent {\bf Remark 1.}

(i) For $0 \leq r \leq n-1$, a code $C \subseteq F^n$ is $r$-separating if, and only if, it is also $(n-r-1)$-separating, because $B_r(x)=F^n \setminus B_{n-r-1}(x+1^n)$ for all $x\in F^n$.

(ii) Since a separating code is such that at most one vertex can be covered by zero codeword, the size of an optimum $r$-separating code in~$F^n$ is $M_r(n)$ or $M_r(n)-1$, and we have:
\begin{equation} \label{separat}
M_{\max\{r,n-r-1\}}(n) \leq M_{\min\{r,n-r-1\}}(n) \leq M_{\max\{r,n-r-1\}}(n) +1,
\end{equation}
i.e., the symmetry, with respect to $\lfloor (n-1)/2 \rfloor$, observed for separating codes, still holds, within one, for identifying codes.
\begin{lemma}\label{lemm}
For all $p\geq 1$ and $\Delta \in \{0, 1, \ldots, p-1\}$, the set $F^p\setminus \{0^p\}$ is $\Delta$-separating. \qed
\end{lemma}
The following theorem is inspired by \cite[Th.~9]{karp98} and \cite[Ex.~2 and Th.~4]{exoo08}. Starting with an $(r,n)$ code~$C$, we intend to see how the direct sum $C\oplus F^p$ can be used for constructing an $(r,n+p)$ code. In construction~${\cal C}$2, $k$ is an additional parameter on which we can act.

More comments on how to understand and use this theorem are given after its statement.
\begin{theorem} \label{hallucin}
Let $r\geq 1$, $p\geq 1$, and $k\in\{0,1, \ldots, p-1\}$; let $C$ be an $(r,n)$ code and 
%\begin{equation} \label{setX}
$$X_p=\{ x\in F^n: \forall c\in C, d(x,c)\leq r-p \; \; or \; \; d(x,c)>r\}.$$
%\end{equation}
\indent Construction ${\cal C}$1: Let $Y_p \subseteq F^n$ be a (minimum) set such that for every $x\in X_p$ there exists $y\in Y_{p}$ with $r-p+1 \leq d(x,y) \leq r$. Then
\begin{equation} \label{opla}
C' = \big( C\oplus F^p \big) \cup \big( Y_{p} \oplus (F^p \setminus \{0^p\}) \big)
%C'=( C \cup Y_p) \oplus F^p
\end{equation}
is $(r,n+p)$. 

Construction ${\cal C}$2: Let $Y_{p,k}\subseteq F^n$ be a (minimum) set such that for every $x\in X_p$ there exists $y\in Y_{p,k}$ with $d(x,y)=r-k$, and let $C_{p,k}$ be a (minimum) $k$-separating code in~$F^p$. Then
\begin{equation} \label{oplus}
C'=( C\oplus F^p ) \cup (Y_{p,k} \oplus C_{p,k})
\end{equation}
is $(r,n+p)$.
\end{theorem}
\noindent {\bf Proof.} See the proof of Theorem~\ref{hallucinbis}, which contains Theorem~\ref{hallucin} as a parti\-cular case. \qed

\medskip

\noindent Theorem~\ref{hallucin} calls for several remarks, in order to make its dry technicity more friendly.

\medskip

\noindent {\bf Remark 2.} Ideally, $X_p=\emptyset$; then $C\oplus F^p$ is $(r,n+p)$. This is Th.~4 in~\cite{exoo08} (Th.~1 in~\cite{blas01} for $r=1$). {\it This is the case as soon as} $\; p\geq r+1$, cf. Cor.~3 in~\cite{exoo08} (Th.~2 in~\cite{blas01} for $r=1$). Therefore we can limit ourselves to
$$p\leq r.$$
On the other hand, we have
$$X_1 \supseteq X_2 \supseteq \ldots \supseteq X_r,$$
so the smaller the number~$p$, probably the more difficult to jump to length $n+p$ without having a large set $Y_p$ or~$Y_{p,k}$.

\medskip

\noindent {\bf Remark 3.} In construction~${\cal C}$1, we build a minimum set $Y_p$ using the union of $p$ spheres of radii ranging from $r-p+1$ to~$r$, whereas in construction~${\cal C}$2, for $Y_{p,k}$ we use only one sphere of radius ~$r-k$. We can therefore hope for a set $Y_p$ (much) smaller than each set $Y_{p,k}$. The price to pay is that $|Y_p|$ has to be multiplied by~$2^p -1$, whereas $|Y_{p,k}|$ has a (much) smaller factor.

%When $p=1$, construction~(2) is not worse than construction~(1): $k=0$ and $\{0\}$ is a $0$-separating code in~$F$. But there is only one sphere for the construction of~$Y_1$, and $Y_1=Y_{1,0}$. So (\ref{opla}) yields $|C'|=2|C|+2|Y_1|$, whereas (\ref{oplus}) yields $|C'|=2|C|+|Y_1|$.
When $k=0$ or $k=p-1$, the smallest $k$-separating codes in $F^p$ have size $2^p-1$, and construction~${\cal C}$2 is not better than construction~${\cal C}$1; therefore, for construction~${\cal C}$2 we can limit ourselves to the cases
$$1\leq k \leq p-2, \: \: 3\leq p \leq r.$$
For different values of $p$ and $k$, it seems very difficult to compare constructions~${\cal C}$1 and~${\cal C}$2, or constructions~${\cal C}$2 between themselves. For a fixed~$p$, $k$ varies from $1$ to $p-2$. When $k$ increases, up to $\lfloor (p-1)/2 \rfloor$, it may be that $|Y_{p,k}|$ increases and $|C_{p,k}|$ decreases (and, by Remark~1(i) before Theorem~\ref{hallucin}, in this case $|C_{p,k}|$ would increase when $k$ ranges from $\lfloor (p-1)/2 \rfloor +1$ to~$p-2$); but actually the former hypothesis highly depends on particular situations (see Example~1 below), and the latter, more general, remains to be proved.

\medskip

\noindent {\bf Example 1.} In $F^{10}$, consider the five vectors $x_1=1^2|0^8$, $x_2=0^2|1^2|0^6$, $x_3=0^4|1^2|0^4$, $x_4=0^6|1^2|0^2$, $x_5=0^8|1^2$. Then $0^{10}$ is at distance two from each of them, but it is easy to see %(A BIEN VERIFIER !!!)
that it is impossible to find a vector which is at distance one from each of them or a vector which is at distance three from each of them. So, if $X_p=\{x_1,x_2,x_3,x_4,x_5\}$, then we have $|Y_{p,r}|=5$, $|Y_{p,r-1}|>1$, $|Y_{p,r-2}|=1$ and $|Y_{p,r-3}|>1$.

\medskip

\noindent This could indicate that, in the absence of information on~$|Y_{p,k}|$, a reasonable bet is to take $k=\lfloor (p-1)/2 \rfloor$, assuming that $|C_{p,k}|$ is minimum for this~$k$. Let us give two small examples.

\medskip

\noindent {\bf Example 2.} We use the notation of Theorem~\ref{hallucin}.

-- Case $p=3$; $r\geq 3$, $k=1$.\\
$Y_3$ is such that $d(x,y)=r-2, r-1$ or~$r$, and $|Y_3|$ is multiplied by~$7$.\\ 
%$Y_{3,0}$ is such that $d(x,y)=r$, and $|Y_{3,0}|$ is multiplied by~$7$.\\
$Y_{3,1}$ is such that $d(x,y)=r-1$, and $|Y_{3,1}|$ is multiplied by $M_1(3)-1=3$: $C_{3,1}=\{000,001,100\}$ is $1$-separating in~$F^3$ (but not $1$-identifying: $111$ is not $1$-covered by~$C_{3,1}$).

-- Case $p=5$; $r\geq 5$, $k\in\{1,2,3\}$.\\
$Y_5$: $d(x,y)\in\{r-4, r-3, r-2, r-1,r\}$, and $|Y_5|$ multiplied by~$31$.\\
%$Y_{5,0}$: $d(x,y)=r$, $|Y_{5,0}|$ multiplied by~$31$.\\
$Y_{5,1}$: $d(x,y)=r-1$, $|Y_{5,1}|$ multiplied by $M_1(5)=10$ or by $M_1(5)-1=9$.\\
$Y_{5,2}$: $d(x,y)=r-2$, $|Y_{5,2}|$ multiplied by $M_2(5)=6$ or by $M_2(5)-1=5$.\\ %(ESSAYER DE VOIR SI ON PEUT FAIRE AVEC~5).
$Y_{5,3}$: $d(x,y)=r-3$, $|Y_{5,3}|$ multiplied by $M_1(5)=10$ or by $M_1(5)-1=9$.

\medskip

\noindent {\bf Remark 4.} The definition of $C'$ shows that $|C|$ will have a factor $2^p$, so it seems best, in general, to take a code~$C$ as small as possible. However, it may be that a larger~$C$, together with a (smaller)~$X_p$ inducing a smaller $Y_p$ or $Y_{p,k}$, gives better results.
% NEW:
In practice, since one cannot try everything, we were led to use the best identifying codes at our disposal.

\medskip

\noindent {\bf Open problem.} Among all $(r,n)$ codes $C$ with $|C|=M_r(n)$, is there at least one such that the set $X_r$ defined in Theorem~\ref{hallucin} is empty? If the answer is {\small YES}, then $M_r(n+r) \leq 2^rM_r(n)$; in particular, we would have $M_1(n+1) \leq 2M_1(n)$. Could this be true for $X_p$ for any $p\in\{1,\ldots, r\}$, so that we would have $M_r(n+p) \leq 2^pM_r(n)$?

\medskip

\noindent It is possible to generalize the previous construction, changing both length (from $n$ to $n+p$) and radius (from $r_1$ to $r_1+r_2$), the case $r_2=0$ being exactly Theorem~\ref{hallucin}.
\begin{theorem} \label{hallucinbis}
Let $r_1\geq p\geq r_2 \geq 1$, and $k\in\{0,1, \ldots, p-1\}$; let $C$ be an $(r_1,n)$ code and 
$$X_{p,r_2}=\{x \in F^n: \forall c\in C, d(x,c)\leq r_1-p+r_2 \; \; or \; \; d(x,c)>r_1+r_2\}.$$
\indent Construction ${\cal C}$1: Let $Y_{p,r_2} \subseteq F^n$ be a (minimum) set such that for every $x\in X_{p,r_2}$ there exists $y\in Y_{p,r_2}$ with $r_1-p+r_2+1 \leq d(x,y) \leq r_1+r_2$. Then
$$C' = \big( C\oplus F^p \big) \cup \big( Y_{p,r_2} \oplus (F^p \setminus \{0^p\}) \big)$$
%$$C'=( C \cup Y_{p,r_2}) \oplus F^p$$
is $(r_1+r_2,n+p)$. 

Construction ${\cal C}$2: Let $Y_{p,r_2,k}\subseteq F^n$ be a (minimum) set such that for every $x\in X_{p,r_2}$ there exists $y\in Y_{p,r_2,k}$ with $d(x,y)=r_1+r_2-k$, and let $C_{p,k}$ be a (minimum) $k$-separating code in~$F^p$. Then
$$C'=( C\oplus F^p ) \cup (Y_{p,r_2,k} \oplus C_{p,k})$$
is $(r_1+r_2,n+p)$.
\end{theorem}
\noindent {\bf Proof.} First, we prove, in both constructions, ${\cal C}$1~and~${\cal C}$2, that any $x\in F^{n+p}$ is $(r_1+r_2)$-covered by a codeword in~$C'$. We write $x=x_1|x_2$ with $x_1\in F^n$, $x_2\in F^p$. Because $C$ is $r_1$-identifying in~$F^n$, there is a codeword $c\in C$ such that $d(c,x_1)\leq r_1$. Therefore, $d(c|x_2,x_1|x_2)\leq r_1 \leq r_1+r_2$, with $c|x_2 \in C\oplus F^p \subseteq C'$.

Next, we prove that, given any two vectors $x,y\in F^{n+p}$ ($x\neq y$), there is a codeword in~$C'$ which $(r_1+r_2)$-separates them. We write $x=x_1|x_2$, $y=y_1|y_2$, with $x_1,y_1\in F^n$, $x_2,y_2\in F^p$. We distinguish between four cases. The first three cases, (i)---(iii), work for both constructions~${\cal C}$1 and~${\cal C}$2, because only $C\oplus F^p$ is needed.

(i) $x_1\neq y_1$, $x_2 \neq y_2$. Then there is a codeword $c\in C$ such that, say, $d(c,x_1)\leq r_1$ and $d(c,y_1)>r_1$. If $r_2\leq p-1$, then two spheres with radius~$r_2$ and distinct centres are different in~$F^p$, and one is not included in the other. So there is a vector $v\in F^p$ which is within distance~$r_2$ from~$x_2$ and not from~$y_2$. If $r_2=p$, we take $v=y_2+1^p$, so that $d(v,y_2)=r_2$ and $d(v,x_2)\leq r_2$.

In both cases, $d(c|v,x_1|x_2)\leq r_1+r_2$ and $d(c|v,y_1|y_2)>r_1+r_2$, with $c|v \in C\oplus F^p \subseteq C'$.

(ii) $x_1\neq y_1$, $x_2=y_2$. Apply the argument in~(i) with $v=x_2+1^{r_2}|0^{p-r_2}$.

(iii) $x_2 \neq y_2$ and $x_1=y_1 \notin X_{p,r_2}$. Then there is a codeword $c\in C$ such that $r_1-p+r_2+1\leq d(c,x_1)\leq r_1+r_2$. If we set $\Delta=r_1+r_2-d(c,x_1)$, we see that $0\leq \Delta \leq p-1$. Therefore, as in case~(i), we can find a vector $v\in F^p$ which is within distance~$\Delta$ from~$x_2$ and not from~$y_2$. Now $d(c|v,x_1|x_2) \leq d(c,x_1)+\Delta =r_1+r_2$ and $d(c|v,x_1|y_2) >d(c,x_1)+\Delta =r_1+r_2$, with $c|v \in C\oplus F^p \subseteq C'$.

(iv) $x_2 \neq y_2$ and $x_1=y_1 \in X_{p,r_2}$.

In construction~${\cal C}$1, there is a vector $z\in Y_{p,r_2}$ such that $r_1-p+r_2+1 \leq d(z,x_1) \leq r_1+r_2$. Then if we set $\Delta=r_1+r_2-d(z,x_1)$, we see that $0\leq \Delta \leq p-1$, and by Lemma~\ref{lemm}, there is a vector $v\in F^p \setminus \{0^p\}$ which is within distance~$\Delta$ from~$x_2$ and not from~$y_2$, or the other way round. Then $d(z|v,x_1|x_2) \leq d(z,x_1)+\Delta =r_1+r_2$ and $d(z|v,x_1|y_2) >d(z,x_1)+\Delta =r_1+r_2$, or the other way round, with $z|v \in Y_{p,r_2} \oplus (F^p \setminus \{0^p\}) \subseteq C'$, and we have proved that $x$ and $y$ are ($r_1+r_2$)-separated by~$C'$.

In construction~${\cal C}$2, there is a vector $z\in Y_{p,r_2,k}$ such that $d(z,x_1)=r_1+r_2-k$ and a codeword $c \in C_{p,k}$ such that, say, $d(c,x_2) \leq k$ and $d(c,y_2)>k$. Then $d(z|c,x_1|x_2)\leq r_1+r_2$ and $d(z|c,x_1|y_2)> r_1+r_2$, with $z|c \in Y_{p,r_2,k} \oplus C_{p,k} \subseteq C'$. \qed
\subsection{Heuristics : noising and greedy}
\label{sec32}
We have mentioned at the end of Section~\ref{sec2} a result on complexity which suggests that constructing good identifying or discriminating codes in the Hamming space might be hard.

Here, we use two different heuristic methods in order to build good identifying codes from scratch, {\it noising} and {\it greedy}.

Noising algorithms have already been used in~\cite{char02} for the construction of identifying codes in various grids; they constitute a family of metaheuristics, of which one is a generalization of simulated annealing~\cite{char01}. Another of these consists of the following. Once $r$, $n$ and a number of codewords, $c$, have been fixed, we consider codes $C\subseteq F^n$ with $c$ codewords, and we define $NC(C)$ as the number of vectors which are not $r$-covered by~$C$, $NS(C)$ as the number of pairs of vectors not $r$-separated by~$C$, and the {\it evaluation function}
$$f(C) = NC(C) + NS(C),$$
which we try to make equal to zero. An initial random code is chosen, which will be the current code~$C$. We iteratively modify the current code, using an {\it elementary transformation} which consists in replacing a codeword by a noncodeword, thus keeping $|C|=c$.

Now when do we accept an elementary transformation? We cyclically go through all codewords: after looking into the last codeword, we start again with the first one. Looking into a codeword $m$ means that we go through all vectors $s$ in $F^n \setminus C$, we note $C_{m,s}=C\setminus \{m\} \cup \{s\}$, and we compute
$$\Delta (C,m,s)=f(C_{m,s})-f(C).$$
For each $s$, we also compute a noised value
$$\Delta_{\rm noise}(C,m,s)=\Delta (C,m,s)+({\rm \rho} \times {\rm ln}(R)),$$
where $\rho$ is a tuning parameter which we make decrease, and $R$ is a number which is randomly chosen for each new elementary transformation (see below for more details). 

If there is a vector $s$ for which $\Delta (C,m,s)<0$, then we keep a vector $s_0$ which minimizes $\Delta (C,m,s)$.

If for all vectors $s$, we have $\Delta (C,m,s)\geq 0$, then we look for a vector $s_0$ which minimizes $\Delta_{\rm noise} (C,m,s)$, and we keep $s_0$ only if $\Delta_{\rm noise} (C,m,s_0)<0$.

If a vector $s_0$ has been found in one of the two cases above, then we apply the elementary transformation with $C,m$ and~$s_0$, so that $C$ becomes $C\setminus \{m\}$ $\cup \, \{s_0\}$. Otherwise, the current code is not modified after looking into~$m$. After each accepted elementary transformation, we check the evaluation function of the current code: if $f(C)=0$, then $C$ is $r$-identifying.

If we have found an identifying code, we reinitialize the process by removing from the current code~$C$ a codeword $m$ which minimizes $f(C\setminus \{m\})$, and we cyclically go through the remaining codewords.

The parameter $R$ is a real number, randomly chosen, in a uniform way, between zero and one; the noising rate $\rho$ is a positive real number which we decrease arithmetically from an initial value down to zero, and for each value of~$\rho$, we cyclically go through the codewords a certain number of times.

\medskip

\noindent Greedy algorithms are based on the following simple idea: starting from an empty code~$C$, at each step we choose to add in~$C$ a codeword $m$ which will maximize $f(C)-f(C\cup \{m\})$. In case of a tie, the choice is made at random.
\section{Results}
\label{resu}
We give tables of lower and upper bounds on $M_r(n)$ for $1\leq r\leq 5$, $1\leq n \leq 21$. There are boldface figures when the exact value is known. Up to now, the most extensive tables ($r=1$, $n\leq 19$, and $r=2$, $n\leq 21$) had been given in~\cite{exoo08}.
\subsection{Using heuristics}
The upper bounds which are marked by a star in our Tables were obtained by noising methods, whereas a double star indicates a result obtained by a greedy algorithm. For instance, the code consisting of the length-9 binary expressions of the following 114 integers

\medskip

{\small
\begin{tabular}{|c|c|c|c|c|c|c|c|c|c|c|c|} \hline
0 & 1 & 8 & 14 & 17 & 20 & 23 & 29 & 31 & 32 & 37 & 39 \\
45 & 49 & 58 & 59 & 70 & 72 & 73 & 75 & 79 & 82 & 84 & 99 \\
101 & 118 & 120 & 121 & 122 & 126 & 129 & 131 & 139 & 140 & 142 & 148 \\
154 & 157 & 172 & 177 & 182 & 183 & 186 & 188 & 194 & 209 & 215 & 216 \\
219 & 222 & 226 & 227 & 228 & 233 & 239 & 240 & 247 & 263 & 264 & 267 \\ 
268 & 274 & 276 & 295 & 297 & 300 & 306 & 314 & 317 & 319 & 323 & 325 \\ 
339 & 344 & 348 & 350 & 352 & 358 & 364 & 367 & 368 & 369 & 374 & 383 \\
391 & 393 & 395 & 404 & 405 & 406 & 409 & 414 & 416 & 418 & 420 & 425 \\
435 & 440 & 448 & 452 & 453 & 458 & 461 & 467 & 475 & 485 & 489 & 490 \\
494 & 495 & 499 & 508 & 509 & 510 & & & & & & \\ \hline
\end{tabular}
}

\medskip

\noindent is a $(1,9)114$ code obtained by noising. All our best codes can be found, in the same form, at 

{http://www.infres.enst.fr/$\sim$charon/identifyingNcube.html}
\subsection{Applying Theorem \ref{hallucin}}
\label{subapplyTH3}
As more or less direct consequences of the results obtained by noising and greedy methods, we also obtain the following results --- note that the various sets $Y_i, Y_{i,j}$ below are obtained via a greedy-type algorithm.

\medskip

%\noindent (1) Using \cite[ Cor.~2]{exoo08} which states that $M_2(n_1+n_2) \leq M_1(n_1) \cdot M_1(n_2)$:
%\begin{equation} \label{eqq1}
%M_2(21) \leq M_1(10) \cdot M_1(11) \leq 211 \cdot 352 = 74272.
%\end{equation}
%(2) 
(1) Using \cite[Cor.~3]{exoo08} (\cite[Th.~2]{blas01} for $r=1$), mentioned in Remark~2:
%%\begin{equation} \label{eqq2}
%%M_1(17) \leq 4 M_1(15) \leq %4 \cdot 4848 = 
%%19392; \; \; M_1(18) \leq 8 M_1(15) \leq %8 \cdot 4848 = 
%%38784;
%%\end{equation}
\begin{equation} \label{eqq3}
M_1(21) \leq 4 M_1(19) \leq %4 \cdot 65536 = 
262144.
\end{equation}
\begin{equation} \label{eqq4}
M_2(19) \leq 8 M_2(16) \leq %8 \cdot 1858 = 
14864; \; \; M_2(20) \leq 16 M_2(16) \leq %16 \cdot 1858 = 
29728;
\end{equation}
\begin{equation} \label{eqq5}
M_2(21) \leq 32 M_2(16) \leq %32 \cdot 1858 = 
59456.
\end{equation}
\begin{equation} \label{eqq6}
M_3(18) \leq 16 M_3(14) \leq %16 \cdot 181 = 
2896; \; \; M_3(19) \leq 32 M_3(14) \leq %32 \cdot 181 = 
5792;
\end{equation}
\begin{equation} \label{eqq7}
M_3(20) \leq 64 M_3(14) \leq %64 \cdot 181 = 
11584; \; \; M_3(21) \leq 128 M_3(14) \leq %128 \cdot 181 = 
23168.
\end{equation}
\begin{equation} \label{eqq8}
M_4(19) \leq 32 M_4(14) \leq %32 \cdot 76 = 
2432; \; \; M_4(20) \leq 64 M_4(14) \leq %64 \cdot 76 = 
4864;
\end{equation}
\begin{equation} \label{eqq9}
M_4(21) \leq 128 M_4(14) \leq %128 \cdot 76 = 
9728.
\end{equation}
\begin{equation} \label{eqq10}
M_5(19) \leq 64 M_5(13) \leq %64 \cdot 28 = 
1792; \; \; M_5(20) \leq 128 M_5(13) \leq %128 \cdot 28 = 
3584;
\end{equation}
\begin{equation} \label{eqq11}
M_5(21) \leq 256 M_5(13) \leq %256 \cdot 28 = 
7168.
\end{equation}
%(3a) 
(2a) Because we have a $(1,13)1322$ code with $X_1=\emptyset$, we have
\begin{equation} \label{eqq12}
M_1(14) \leq 2 \cdot 1322 = 2644.
\end{equation}
(2b) We have a $(1,15)4848$ code with $|X_1|=128$; unfortunately, because of the distance distribution in~$X_1$, it is impossible to obtain a set $Y_{1}$ with fewer than 128 elements, and therefore, by cons\-truction~${\cal C}$1:
\begin{equation} \label{eqq13}
M_1(16)\leq 2 \cdot 4848 + 128 = 9824.
\end{equation}
(2c) Because the $(1,19)65536$ code from~\cite{exoo08} is such that every vector is $1$-covered by at least two codewords, we have $X_1=\emptyset$ and 
\begin{equation} \label{eqq14}
M_1(20) \leq 2 \cdot 65536 = 131072.
\end{equation}
(2d) We have a $(2,16)1858$ code with $|X_1|=441$ and we found a corres\-ponding set $Y_{1}$ with 151 elements; therefore, by construction~${\cal C}$1:
\begin{equation} \label{eqq15}
M_2(17)\leq 2 \cdot 1858 + 151 = 3867.
\end{equation}
The same $(2,16)1858$ code has $|X_2|=283$, with $|Y_2|=105$, consequently:
\begin{equation} \label{eqq16}
M_2(18) \leq 4 \cdot 1858 + 105 \cdot 3 = 7747.
\end{equation}
(2e) We have a $(3,14)181$ code with $|X_1|=60$ and a set $Y_{1}$ with 13 elements; therefore,
\begin{equation} \label{eqq17}
M_3(15) \leq 2 \cdot 181 + 13 = 375.
\end{equation}
This $(3,14)181$ code has $|X_2|=6$, a set $Y_{2}$ with 4 elements, and we obtain:
\begin{equation} \label{eqq18}
M_3(16) \leq 4 \cdot 181 + 4 \cdot 3 = 736.
\end{equation}
The same $(3,14)181$ code has $X_3=\emptyset$, and
\begin{equation} \label{eqq19}
M_3(17) \leq 8 \cdot 181 = 1448.
\end{equation}
(2f) We have a $(4,14)76$ code with $|X_1|=26$, $|Y_{1}|=4$, yielding
\begin{equation} \label{eqq20}
M_4(15) \leq 2 \cdot 76 + 4 = 156.
\end{equation}
The same $(4,14)76$ code has $X_2=X_3=X_4=\{7577, 8802\}$; these two numbers represent two length-14 vectors at distance 13 from one another, so all the sets $Y_i, Y_{i,j}$ have size two for $i=2,3,4$. In particular, $|Y_{2}|=|Y_{3,1}|=|Y_{4,1}|=2$; therefore, by construction~${\cal C}$1:
\begin{equation} \label{eqq21}
M_4(16) \leq 4 \cdot 76+2\cdot 3 = 310,
\end{equation}
and by construction~${\cal C}$2:
\begin{equation} \label{eqq215}
M_4(17) \leq 8 \cdot 76+2\cdot 3 = 614, \; \; M_4(18) \leq 16 \cdot 76+2\cdot 6 = 1228,
\end{equation}
because optimum $1$-separating codes have size three in~$F^3$ (see Example~2) and have size six in~$F^4$ --- this comes from Remark~1(ii) on separating codes and the fact that $M_1(4)=7$ and $M_2(4)=6$ (see Tables~1 and~2).

\medskip

\noindent (2g) We have a $(5,13)28$ code with $|X_1|=43$, $|Y_{1}|=4$, yielding
\begin{equation} \label{eqq22}
M_5(14) \leq 2 \cdot 28 + 4 = 60.
\end{equation}
This $(5,13)28$ code has $|X_2|=1$, $|Y_{2}|=1$, and therefore
\begin{equation} \label{eqq23}
M_5(15) \leq 4 \cdot 28 + 1 \cdot 3 = 115.
\end{equation}
This same $(5,13)28$ code has $X_3=X_4=X_5=\emptyset$, and so:
\begin{equation} \label{eqq24}
M_5(16) \leq 8 \cdot 28 = 224, \: M_5(17) \leq 16 \cdot 28 = 448, \: M_5(18) \leq 32 \cdot 28 = 896.
\end{equation}
All the nonempty sets $X_i$, $Y_i$, $Y_{i,j}$ mentioned above are given at

{http://www.infres.enst.fr/$\sim$charon/identifyingNcube.html}
\subsection{Further improvements: removing codewords}
\label{remove}
Perhaps Theorems~\ref{hallucin} and~\ref{hallucinbis} can be sharpened, since in practice we observe (with the help of a computer) that the sizes of several %(many? most? all?) 
codes obtained by Theorem~\ref{hallucin} can be reduced by simply removing some of their codewords, which are ``useless''. 
%NEW:
This can also be done with two %%(some) 
of the codes obtained by a greedy algorithm.

As a consequence, we have new upper bounds for some %(?) 
values of $n$ and~$r$, which are marked by a triple star in the Tables. The corresponding codes can be found at

{http://www.infres.enst.fr/$\sim$charon/identifyingNcube.html}\\
%\\
%NEW:
We can observe that when $r$ increases, the reductions can be drastic --- almost 50\% in the case $r=5$, $n=18\,$!
\subsection{Re-applying Theorem \ref{hallucin}}
\label{reuse}
We can again use Theorem~\ref{hallucin} with the newly improved codes obtained in Section~\ref{remove}. 
% Due to time and space limitations, etc
%% a mettre juste avant \subsection{Tables} (4.5)
%%% Due to limitations of time and space, we could not try to remove codewords from these new codes.

\medskip

\noindent (3a) We have a $(2,20)29346$ code with $X_1=\emptyset$, and so
\begin{equation} \label{eqq204}
M_2(21) \leq 2 \cdot 29346 = 58692.
\end{equation}
(3b) We have a $(3,19)5532$ code with $X_1=X_2=\emptyset$, and so
\begin{equation} \label{eqq1515}
M_3(20) \leq 2 \cdot 5532 = 11064, \: M_3(21) \leq 4 \cdot 5532 = 22128.
\end{equation}
(3c) We have a $(4,18)1045$ code with $|X_1|=2$, $|Y_{1}|=2$, yielding
\begin{equation} \label{eqq2201}
M_4(19) \leq 2 \cdot 1045 + 2 = 2092.
\end{equation}
This $(4,18)1045$ code has $X_2=X_3=\emptyset$ and therefore
\begin{equation} \label{eqq2301}
M_4(20) \leq 4 \cdot 1045 = 4180, \: M_4(21) \leq 8 \cdot 1045 = 8360.
\end{equation}
(3d) We have a $(5,18)454$ code with $|X_1|=1$, $|Y_{1}|=1$, yielding
\begin{equation} \label{eqq220}
M_5(19) \leq 2 \cdot 454 + 1 = 909.
\end{equation}
This $(5,18)454$ code has $|X_2|=1$, $|Y_{2}|=1$, and therefore
\begin{equation} \label{eqq230}
M_5(20) \leq 4 \cdot 454 + 1 \cdot 3 = 1819.
\end{equation}
This same $(5,18)454$ code has $X_3=\emptyset$, and so:
\begin{equation} \label{eqq240}
M_5(21) \leq 8 \cdot 454 = 3632.
\end{equation}
%MOVED HERE:
Due to time and space limitations, we could not try to remove codewords from these new codes.
\subsection{Tables}
We give our results for $1\leq r \leq 5$, $r+1\leq n \leq 21$. For some values of $r$ and~$n$, we give two upper bounds, the first one from Section~\ref{subapplyTH3}, the second one from Sections~\ref{remove} or~\ref{reuse}, so that one can see how we used Theorem~\ref{hallucin} then possibly removed codewords and possibly reused Theorem~\ref{hallucin}.

We think that there is still room for ameliorations, and we encourage the reader to improve on these upper bounds.

\bigskip

\begin{tabular}{ll}
\multicolumn{1}{l}{Key to Tables}\\ \hline
Lower bounds & Upper bounds\\ \hline
a \cite[Th. 1(iii)]{karp98} & A \cite{karp98}\\
b \cite[Th. 2]{karp98} & B $M_{n-1}(n)=2^{n}-1$ \cite[Th. 5]{blas00}\\
c \cite[Th. 3]{karp98} & C \cite[Th. 4]{blas01}\\
d \cite[Th. 4]{blas01} & D \cite[Th. 5]{blas01}\\
e \cite[Th. 11]{blas01} & E \cite[Th. 6]{blas01}\\
f $M_{n-1}(n)=2^{n}-1$ \cite[Th. 5]{blas00} $\; \; \; \; \; \; \; \;$ & F \cite[Tables 3 and 4]{exoo08}\\
g \cite[Th. 6]{blas00} & G \cite{exoo99}\\
h \cite[Table 4]{exoo08} & H \cite[Th. 6]{blas00}\\ 
i \cite[Cor. 4]{laih07} & $^*$ {\it noising} \\
j \cite[Cor. 5]{laih07} & $^{**}$ {\it greedy }\\
k \cite[Cor. 7]{laih07} & $^{***}$ {\it removing} codewords\\
$\ell$ by (\ref{separat}) and $M_1(5)=10$ & (x) inequality (x)\\ 
m \cite[Table 3.1]{rant07} & \\ \hline
\end{tabular}

\bigskip

{\small
\begin{tabular}{c||c|c|c|}
$n$ & lower bound & 1st upper bound & 2nd upper bound \\ \hline \hline
{\bf 2} & a {\bf 3} & {\bf 3} B & \\
{\bf 3} & b {\bf 4} & {\bf 4} A & \\
{\bf 4} & d {\bf 7} & {\bf 7} C & \\
{\bf 5} & b {\bf 10}& {\bf 10} A & \\ \hline
6 & c 18 & 19 D & \\
{\bf 7} & e {\bf 32} & {\bf 32} E & \\
8 & c 56 & 62 G,F & \\
9 & c 101 & $114^*$ & \\
10 & c 183 & $211^*$ & \\ \hline
11 & c 337 & 352 F & \\
12 & c 623 & $688^*$ & \\
13 & c 1158 & $1322^*$ & \\
14 & c 2164 & $2644$ (\ref{eqq12}) & \\
15 & c 4063 & $4848^*$ & \\ \hline
16 & c 7654 & $9824$ (\ref{eqq13}) & 9779$^{***}$ \\
17 & c 14469 & $19043^{**}$ & 19026$^{***}$ \\
18 & c 27434 & $36423^{**}$ & 36406$^{***}$ \\
19 & c 52155 & 65536 F & \\
20 & c 99392 & $131072$ (\ref{eqq14}) & \\ \hline
21 & c 189829 & $262144$ (\ref{eqq3}) & \\
\multicolumn{4}{c}{}\\
\multicolumn{4}{c}{Table 1: Lower and upper bounds, $r=1$.}\\
\end{tabular}
}
%%% Table r=1 finie, codes en html tous faits

\bigskip

{\small
\begin{tabular}{c||c|c|c|}
$n$ & lower bound & 1st upper bound & 2nd upper bound \\ \hline \hline
{\bf 3} & f {\bf 7} & {\bf 7} B & \\
{\bf 4} & g {\bf 6} & {\bf 6} H & \\
{\bf 5} & a {\bf 6} & {\bf 6} H & \\ \hline
{\bf 6} & a {\bf 8} & {\bf 8} H & \\
{\bf 7} & h {\bf 14} & {\bf 14} F & \\
8 & h 20 & 21 F & \\
9 & m 26 & 32$^{*}$ & \\
10 & i 41 & $60^*$ & \\ \hline
11 & i 67 & $106^{**}$ & \\
12 & i 112 & $185^{**}$ & \\
13 & i 190 & $328^{**}$ & \\
14 & i 326 & $580^{**}$ & \\
15 & i 567 & $1032^{**}$ & \\ \hline
16 & i 995 & $1858^{**}$ & \\
17 & i 1761 & $3867$ (\ref{eqq15}) & 3785$^{***}$ \\
18 & i 3141 & $7747$ (\ref{eqq16}) & 7609$^{***}$ \\
19 & i 5638 & $14864$ (\ref{eqq4}) & 14673$^{***}$ \\
20 & i 10179 & $29728$ (\ref{eqq4}) & 29346$^{***}$ \\ \hline
21 & i 18471 & $59456$ (\ref{eqq5}) & 58692 (\ref{eqq204}) \\
\multicolumn{4}{c}{}\\
\multicolumn{4}{c}{Table 2: Lower and upper bounds, $r=2$.}\\
\end{tabular}
}

\bigskip

{\small
\begin{tabular}{c||c|c|c|}
$n$ & lower bound & 1st upper bound & 2nd upper bound \\ \hline \hline
{\bf 4} & f {\bf 15} & {\bf 15} B & \\
5 & $\ell$ 9 & $10^*$ & \\ \hline
{\bf 6} & a {\bf 7} & {\bf 7}$^*$ & \\
{\bf 7} & a {\bf 8} & {\bf 8}$^*$ & \\
8 & a 10 & $13^*$ & \\
9 & a 13 & $17^*$ & \\
10 & a 18 & $28^*$ & \\ \hline
11 & a 25 & $37^{**}$ & \\
12 & a 39 & $68^{**}$ & \\
13 & a 61 & $112^{**}$ & \\
14 & a 95 & $181^{**}$ & \\
15 & a 151 & $375$ (\ref{eqq17}) & 356$^{***}$ \\ \hline
16 & a 241 & $736$ (\ref{eqq18}) & 700$^{***}$ \\
17 & a 383 & $1448$ (\ref{eqq19})& 1387$^{***}$ \\
18 & a 608 & $2896$ (\ref{eqq6}) & 2766$^{***}$ \\
19 & a 959 & $5792$ (\ref{eqq6}) & 5532$^{***}$ \\
20 & k 1593 & $11584$ (\ref{eqq7}) & $11064$ (\ref{eqq1515}) \\ \hline
21 & j 2722 & $23168$ (\ref{eqq7}) & $22128$ (\ref{eqq1515}) \\ 
\multicolumn{4}{c}{}\\
\multicolumn{4}{c}{Table 3: Lower and upper bounds, $r=3$.}\\
\end{tabular}
}

\bigskip

{\small
\begin{tabular}{c||c|c|c|}
$n$ & lower bound & 1st upper bound & 2nd upper bound \\ \hline \hline
{\bf 5} & f {\bf 31} & {\bf 31} B & \\ \hline
6 & a 7 & $18^*$ & \\
7 & a 8 & $14^*$ & \\
8 & a 9 & $13^*$ & \\
9 & a 10 & $14^*$ & \\
10 & a 12 & $16^*$ & \\ \hline
11 & a 15 & $20^*$ & \\
12 & a 19 & $34^{**}$ & \\
13 & a 27 & $48^{**}$ & \\
14 & a 38 & $76^{**}$ & \\
15 & a 54 & $156$ (\ref{eqq20}) & 142$^{***}$ \\ \hline
16 & a 77 & $310$ (\ref{eqq21}) & 272$^{***}$ \\
17 & a 121 & $614$ (\ref{eqq215}) & 530$^{***}$ \\
18 & a 190 & $1228$ (\ref{eqq215}) & 1045$^{***}$\\
19 & a 304 & $2432$ (\ref{eqq8}) & $2092$ (\ref{eqq2201}) \\
20 & a 489 & $4864$ (\ref{eqq8}) & $4180$ (\ref{eqq2301}) \\ \hline
21 & a 792 & $9728$ (\ref{eqq9}) & $8360$ (\ref{eqq2301}) \\ 
\multicolumn{4}{c}{}\\
\multicolumn{4}{c}{Table 4: Lower and upper bounds, $r=4$.}\\
\end{tabular}
}

\bigskip

{\small
\begin{tabular}{c||c|c|c|}
$n$ & lower bound & 1st upper bound & 2nd upper bound \\ \hline \hline
{\bf 6} & f {\bf 63} & {\bf 63} B & \\
7 & a 8 & $35^*$ & \\
8 & a 9 & $22^*$ & \\
9 & a 10 & $17^*$ & \\
10 & a 11 & $19^*$ & \\ \hline
11 & a 12 & $19^{**}$ & \\
12 & a 14 & $25^{**}$ & \\
13 & a 17 & $28^{**}$ & \\
14 & a 21 & 60 (\ref{eqq22}) & 48$^{***}$ \\
15 & a 28 & $115$ (\ref{eqq23}) & 75$^{***}$ \\ \hline
16 & a 37 & $224$ (\ref{eqq24}) & 127$^{***}$ \\
17 & a 53 & $448$ (\ref{eqq24}) & 232$^{***}$ \\
18 & a 77 & $896$ (\ref{eqq24}) & 454$^{***}$ \\
19 & a 112 & $1792$ (\ref{eqq10}) & 909 (\ref{eqq220}) \\
20 & a 161 & $3584$ (\ref{eqq10}) & 1819 (\ref{eqq230}) \\ \hline
21 & a 229 & $7168$ (\ref{eqq11}) & 3632 (\ref{eqq240}) \\ 
\multicolumn{4}{c}{}\\
\multicolumn{4}{c}{Table 5: Lower and upper bounds, $r=5$.}\\
\end{tabular}
}

%%% Table r=5 finie, codes en html tous faits
\subsection{Conclusion}
By mixing both heuristic and theoretical constructing arguments, we were able to present numerous upper bounds on $M_r(n)$, the smallest possible cardinality of an $r$-identifying code in~$F^n$: we first used heuristics for constructions of codes from scratch, we then used some of these codes to build new codes with the help of Theorem~\ref{hallucin}; after that, the computer possibly removed codewords from these codes, and eventually we reapplied Theorem~\ref{hallucin}.

There still remains a large, challenging gap between the lower and upper bounds for most of the values of $r,n$ in Tables~1--5.

Of course, all these bounds are transposable to discriminating codes, by Corollary~\ref{coroh}.

\bigskip

\noindent {\bf Acknowledgment}

\medskip

\noindent Our thanks to Geoffrey Exoo, Tero Laihonen and Sanna Ranto, who let us use~\cite{exoo08} and~\cite{laih07} at a time when they were only submitted papers.
\label{sec33}

\end{document}